\documentclass[nofootinbib,prx,aps,superscriptaddress,notitlepage,reprint,preprintnumbers,showpacs,showkeys,longbibliography]{revtex4-2}
\usepackage{latexsym,graphicx,amssymb,amsmath,mathrsfs,color}
\usepackage[utf8]{inputenc}
\usepackage{setspace,bm}
\usepackage[normalem]{ulem}

\DeclareSymbolFont{bbold}{U}{bbold}{m}{n}
\DeclareSymbolFontAlphabet{\mathbbold}{bbold}

\newcommand{\be}{\begin{equation}}      
\newcommand{\ee}{\end{equation}}      
\newcommand{\bea}{\begin{eqnarray}}      
\newcommand{\eea}{\end{eqnarray}}

\newcommand{\Tr}{\,\textrm{Tr}\,}
\newcommand{\Log}{\,\textrm{Log}\,}
\newcommand{\const}{\,\textrm{const.}\,}
\newcommand{\sccon}{\,\textrm{purely scalar contributions}\,}

\makeatletter
\renewcommand\appendix{\par
\setcounter{section}{0}
\setcounter{subsection}{0}
\gdef\thesection{\appendixname\space\@Alph\c@section}}

\long\def\unmarkedfootnote#1{{\long\def\@makefntext##1{##1}\footnotetext{#1}}}
\makeatother

\begin{document} 
\title{Global Fixed Point Potentials in the Abelian Higgs Model with $N$ flavors}

\author{Gergely Fej\H{o}s}
\email{gergely.fejos@ttk.elte.hu}
\affiliation{Institute of Physics and Astronomy, E\"otv\"os University, 1117 Budapest, Hungary}
\affiliation{Physics Department, Brookhaven National Laboratory, Upton, New York 11973, USA}
\affiliation{RIKEN Center for Interdisciplinary Theoretical and Mathematical Sciences (iTHEMS), Wako, Saitama 351-0198, Japan}
\author{Shunsuke Yabunaka}
\email{yabunaka123@gmail.com}
\affiliation{Advanced Science Research Center, Japan Atomic Energy Agency, Tokai, 319-1195, Japan}

\begin{abstract}
Existence of charged fixed points in the Abelian Higgs model with $N$ flavors in $d$ dimensions is studied using the functional renormalization group. We numerically solve the coupled fixed point equations for the scale dependent charge and the non-perturbative effective potential for the scalar field. We show that the $\epsilon=4-d$ expansion, famously successful in theories with $O(N)$ symmetry, fails to produce reliable results when taking the $\epsilon \rightarrow 1$ limit. By determining global fixed point potentials, it is shown that the critical flavor number at which charged fixed points appear, modifies significantly compared to the perturbative treatment. In $d=3$, signs of a richer fixed point structure with presumably multicritical fixed points are also found. Discussions include subtleties of the gauge fixing and the corresponding modified Ward-Takahashi identities, including the possibility of a nonzero dimensionless photon mass at the infrared fixed point.
\end{abstract}

\maketitle

\section{Introduction}

The $N$-component Abelian Higgs model is one of the natural generalizations of the $O(2N)$ symmetric scalar theory, where the dynamics of an $N$-component complex scalar is extended with a $U(1)$ dynamical gauge field. The model has been extensively studied in the past as it is the simplest theory that combines gauge invariance with spontaneous symmetry breaking leading to massive gauge bosons. It provides a unified framework not only for the Higgs mechanism in particle physics, but also describes important properties of superconductivity in condensed matter physics (Meissner-effect, flux quantization, vortex dynamics, etc.) \cite{degennes}. The Abelian Higgs model also serves as a particularly simple example demonstrating how radiative corrections can generate spontaneous symmetry breaking, also known as the Coleman-Weinberg mechanism \cite{Coleman-Weinberg}. On top of its continuous formulation, the lattice Abelian Higgs model has also gained significant interest in recent times \cite{Pelissetto:2019thf,Bonati:2023bcu,Bonati2021,Bonati:2024lht}.

The three dimensional, Euclidean formulation of the model with $N=1$, sometimes called the Ginzburg-Landau theory, can be considered as a phenomenological model describing conventional superconductivity. In such a system, it is easy to see that after integrating out the gauge degrees of freedom, through a cubic term arising in the effective potential, the Coleman-Weinberg mechanism essentially predicts that the finite temperature phase transition between the normal and the superconducting phases must always be first-order, irrespective of the self coupling of the dynamical scalar field \cite{halperin1974first}. This result can also seen in perturbative renormalization group (RG) analyses, in particular via the $\epsilon$ expansion around $d=4$. When one extrapolates the resulting $\beta$ functions to $d=3$, the absence of an infrared stable fixed point at the critical temperature provides indirect evidence that a second-order transition cannot occur \cite{Zinn-Justin-book}. In contrast, Monte Carlo analyses have shown a long time ago that for a large enough Ginzburg-Landau parameter\footnote{The Ginzburg-Landau parameter is defined as the ratio between the penetration depth and the correlation length, $\kappa=\lambda/\xi$.}, $\kappa$, i.e., when the self coupling of the scalar field is much stronger than the gauge coupling, the transition becomes second-order, showing the existence of a tricritical point \cite{Mo2002,Bartholomew1983}. The second-order transition in question is conjectured to belong to the three-dimensional XY universality class \cite{Kiometzis1995,Olsson1998}, and one may argue that the origin of criticality is due to vortex fluctuations \cite{Kleinert1982,Kleinert2006}, which are missed in perturbative treatments.

If one returns to generic flavor numbers, the perturbative RG at the leading order shows the existence of the pair of a critical and a bicritical charged fixed point\footnote{This also means the existence of a tricritical point in the phase diagram.}, but only if $N\geq N^{(1)}_c=183$ \cite{halperin1974first}, being very far from $N=1$. This result does not seem to be stable against higher order corrections. Results are available up to four loop order \cite{Kolnberger1990,Ihrig,Huang}, where $N_c$ gets modified as $N^{(2)}_c=-137$, $N_c^{(3)}=9$ and $N_c^{(4)}=75$ at higher orders, respectively. Resummation techniques lead to approximately $N_c \approx 12$ \cite{Ihrig}. The problem was also tackled by a semi-perturbative $d=3$ calculation based on the field-theoretical RG approach \cite{Kleinert2003}, but the discrepancy remained. It should be noted that later it was argued that the fixed point responsible for the transition at $N=1$ is actually not connected analytically to those found for larger $N$ values \cite{Nahum2015}, but RG analyses are yet to provide evidence.

When increasing the component number from $N=1$, the situation becomes more subtle. Utilizing the lattice Abelian-Higgs model, it was found that even though the transition is second order for large enough $\kappa$ at $N=1$, when increasing the component number, it always becomes weakly first order for $1<N<N_c$ \cite{Nahum2015,Bonati2021}, and then can turn back to second order for $N_c<N$, for large enough $\kappa$. The critical value, $N_c$, was conjectured to be in the interval $4<N_c<10$ \cite{Bonati2021}. The perturbative RG, therefore, fails in two ways; it seems to predict $N_c$ to be too large at the leading order (i.e., $N_c^{(1)}=183$) without any sign of convergence, and it cannot capture the second order nature of the transition for a single complex scalar.

The apparent contradiction between the perturbative RG results and that of Monte Carlo simulations can also be addressed using the functional RG (FRG) \cite{Wetterich:1992yh,dupuis2021}. Early studies \cite{bergerhoff1996phase,bergerhoff1996PRB} were promising, and they found a charged critical fixed point for any component number that can correspond to second-order transitions. However, they do not provide convincing evidence of the tricritical behavior in case of a single complex scalar, and there were no signs of a purely first order window in terms of $N$. 

Later, also using the FRG, via the generalization of the $R_\xi$ gauges, tricritical behavior was found for $N=1$ \cite{Fejos2016}. The resulting value for the critical $\kappa$ was found to be close to that of Monte Carlo simulations, however, in contradiction to \cite{Nahum2015}, the corresponding fixed points were analytically connected in the entire $N$ range \cite{Fejos2017}, and there was no first-order window whatsoever. As emphasized in these studies, the existence of a tricritical point for $N=1$ is regularization sensitive, and according to \cite{Fejos2016,Fejos2017}, the only way the fixed point structure leading to a tricritical point survives for $N=1$ is that the regulators are introduced to eigenmodes of the propagator matrix, rather than to the original field variables. Since the former depend on the actual background of the scalar field, strictly speaking, the RG equation for the effective action, also known as the Wetterich equation, needs to be modified, which was not done in Refs. \cite{Fejos2016,Fejos2017}. (We will explain this further throughout the text.) What should be emphasized here is that as a result of the above, findings of the aforementioned studies should be taken with caution.

All the FRG (or RG) related works cited above deal with flows of couplings that were already present in the ultra-violet. That is, the real power of the non-perturbative nature of the FRG was never fully exhausted. Therefore, in this study, we aim to adopt a more conservative, yet straightforward approach. Using the FRG, regulators will be associated to the original field variables, leaving the flow equation intact. We employ the usual covariant gauge fixing in the Local Potential Approximation' (LPA'), where the effective action is approximated with the sum of the standard kinetic terms with wave function renormalizations and a fully non-perturbative potential. The latter will be acquired numerically on a grid, without relying on Taylor expansion. Our goal is, therefore, to explore the existence of global fixed point potentials as a function of $N$ and $d$, which has never been done in the Abelian Higgs model. Our method will also give an estimate on the critical flavor number, $N_c$.

The paper is organized as follows. In Sec. II, we introduce the model, gauge fixing and the corresponding notations. Sec. III is dedicated to the RG flow equations of the potential and the wave function renormalizations. The regulator matrix is also defined here. In Sec. IV, we show derivations of the modified Ward-Takahashi identities, where the role of the gauge mass is also explained. Sec. V is devoted for the numerical results, while Sec. VI contains the summary.

\section{Abelian Higgs model}

In this paper we are investigating the field theory of an $N$-component complex scalar field ($\phi$) with $U(1)$ gauge symmetry. In d-dimensional Euclidean space the classical action is as follows:
\bea
\label{Eq:Scl}
S=\int_x {\cal L}=\int d^dx &&\hspace{-0.4cm}\Big[(D_i \phi^a)^\dagger D_i \phi^a+m^2\phi^{\dagger a}\phi^a \nonumber\\
&&\hspace{-0.2cm}+\frac{\lambda}{6}(\phi^{\dagger a} \phi^a)^2+\frac14 F_{ij} F_{ij} \Big],
\eea
where $D_i=\partial_i - ieA_i$ is the covariant derivative with $A_i$ being the gauge field ($i=1,...d)$, $F_{ij}=\partial_i A_j-\partial_j A_i$ is the $U(1)$ field strength tensor, $\phi^a$ is an $N$-component complex scalar ($a=1,...N$), while $m^2$ and $\lambda>0$ are constants. Gauge symmetry is realized by the following local transformation:
\bea
\label{Eq:trf}
\delta \phi^a(x) = ie\theta(x)\phi^a(x), \quad \delta A_i(x) = \partial_i \theta(x),
\eea
where $\theta(x)$ is an infinitesimal parameter. We choose to work with covariant gauge fixing, which in effect adds a
\bea
\label{Eq:gf}
{\cal L}_{gf}=\frac{1}{2\xi}(\partial_i A_i)^2
\eea
piece to the Lagrangian, where $\xi$ is the gauge fixing parameter.

In this study we employ the FRG formalism, and search for scaling behaviors of the scale dependent effective action, which is defined by leaving out modes with momenta $|q| \lesssim k$ from the ordinary quantum effective action, $\Gamma$, where $k$ plays the role of the scale separation variable. The reader is referred for the details to \cite{Wetterich:1992yh,dupuis2021,Delamotte2012}. By construction, $k$ becomes the parameter of the effective action, $\Gamma \rightarrow \Gamma_k$.

For practical calculations, we need an ansatz for $\Gamma_k$. Here, we employ the so-called LPA' approximation, where the scale dependent effective action takes the form of the classical action, but with $k$-dependent couplings, and we also include wave function renormalizations for both the scalar and the gauge fields. This leads to
\bea
\label{Eq:ansatz}
\Gamma_{k}&\!\!\!=\!\!\!&\int \Big[ Z_{\phi,k}(\hat{D}_i \phi^a)^\dagger(\hat{D}_i \phi^a)+V_k(\rho)\nonumber\\
 &\!\!\!+\!\!\!&\frac{Z_{A,k}}{2}A_{i}\left[-\partial^{2}\delta_{ij}+\left(1-\xi_k^{-1}\right)\partial_{i}\partial_{j}\right]A_{j}+\frac{1}{2}m_{A,k}^{2}A_{i}^{2}\Big]\nonumber\\
&\!\!\!=\!\!\!& \int_{x}\Big[Z_{\phi,k}\partial_{i}\phi^{\dagger}\partial_{i}\phi + V_k(\rho) \nonumber\\
&\!\!\!-\!\!\!&ieZ_{e,k}\left(\partial_{i}\phi^{\dagger}\phi-\partial_{i}\phi\phi^{\dagger}\right)A_{i}+\frac{Z_{e,k}^{2}}{Z_{\phi,k}}e^{2}A_{i}^{2}\phi^{\dagger}\phi\nonumber\\
&\!\!\!+\!\!\!&\frac{Z_{A,k}}{2}A_{i}\left[-\partial^{2}\delta_{ij}+\left(1-\xi_k^{-1}\right)\partial_{i}\partial_{j}\right]A_{j}+\frac{1}{2}m_{A,k}^{2}A_{i}^{2}\Big],\nonumber\\
\eea
where $\hat{D}_i = \partial_i - ieZ_{e,k}/Z_{\phi,k}$ and we defined $\rho=\phi^{\dagger a} \phi^a$. Eq. (\ref{Eq:ansatz}) can be obtained from (\ref{Eq:Scl}) and (\ref{Eq:gf}) by employing the rescalings $\phi^a \rightarrow Z_{\phi,k}^{1/2} \phi^a$, $A_i \rightarrow Z_{A,k}^{1/2}A_i$, $e\rightarrow eZ_{e,k}/(Z_{\phi,k}Z_{A,k}^{1/2})$, while promoting the scalar field dependence to a fully non-perturbative potential, $V_k$. Note that, we also added a scale dependent mass term for the gauge field, whose role will be clarified later. In accordance with the rescaling of $e$, the flowing charge can be defined as $e_k=e Z_{e,k}/(Z_{\phi,k}Z_{A,k}^{1/2})$.

One can conveniently decompose $\phi^a$ into two sets of real fields, $s^a$ and $\pi^a$, as
\bea
\phi^{a}=\frac{s^{a}+i\pi^{a}}{\sqrt{2}},
\eea
which, in the minimum point of the effective action yields $\langle s^a \rangle=v\delta_{a,1}$ and $\langle \pi^{a}\rangle=0$ ($a=1,2...N$).

\section{Flow equations}

The scale dependent quantum effective action obeys the Wetterich equation \cite{Wetterich:1992yh}:
\bea
\label{Eq:wet}
k\partial_k \Gamma_k = \frac12 \Tr \Big[k\partial_k {\cal R}_k (\Gamma_k^{(2)}+{\cal R}_k)^{-1}],
\eea
where $\Gamma_k^{(2)}$ is the second derivative matrix of $\Gamma_k$, while ${\cal R}_k$ is the regulator, responsible for scale separation. Eq. (\ref{Eq:wet}) can also be written as
\bea
\label{Eq:wet2}
k\partial_k \Gamma_k = \frac12 k\tilde{\partial}_k\Tr \Log (\Gamma_k^{(2)}+{\cal R}_k),
\eea
where by definition $\tilde{\partial}_k$ acts only on ${\cal R}_k$. In Eq. (\ref{Eq:wet}) and (\ref{Eq:wet2}) both the trace and the inverse operations need to be taken both in the functional and in the matrix senses. Eq. (\ref{Eq:wet}) serves as a master equation to derive individual flow equations for the scale dependent quantities.

In what follows, we take the $\xi\rightarrow0$ limit, i.e., employ the Landau gauge, as it is known to be a fixed point of the FRG flows \cite{Ellwanger:1995qf,Litim:1998qi}. With this choice the gauge propagator is transverse and the scalar propagator is completely diagonal. As a result, we can associate an optimal regulator function, i.e. $R_k(q)=(k^2-q^2)\Theta(k^2-q^2)$, to each mode in an equal fashion. The regulator matrix is, therefore,
\bea
\label{Eq:reg}
{\cal R}_{A_i,A_j} &=& \delta_{ij}Z_{A,k}R_k(q), \quad {\cal R}_{s_i,s_j} = \delta_{ij}Z_{\phi,k}R_k(q), \nonumber\\
{\cal R}_{\pi_i,\pi_j} &=& \delta_{ij}Z_{\phi,k}R_k(q).
\eea
With these choices, the inverse of the regulated two point function take the following form:
\bea
\label{Eq:regprop}
(\Gamma_k^{(2)}+{\cal R}_k)^{-1}_{A_i,A_j}&=&\Big(\delta_{ij}-\frac{q_i q_j}{q^2}\Big)\frac{1}{Z_{A,k}q_R^2+\frac{Z_{e,k}^2}{Z_{\phi,k}}2e^2\rho}, \nonumber\\
(\Gamma_k^{(2)}+{\cal R}_k)^{-1}_{s_i,s_j}&=&\frac{\delta_{ij}}{Z_{\phi,k}q_R^2+V_k'+2\rho V_k'' \delta_{i1}}, \nonumber\\
(\Gamma_k^{(2)}+{\cal R}_k)^{-1}_{\pi_i,\pi_j}&=&\frac{\delta_{ij}}{Z_{\phi,k}q_R^2+V_k'},
\eea
where $q_R^2 = q^2+R_k(q)$. Note that $\rho=v^2/2$. From here onwards, we impose the constraint $Z_{e,k}=Z_{\phi,k}$ at all scales, which is one of the Ward-Takahashi identities at $k=0$.

At this point we mention that if we did not use the Landau gauge, the propagator matrix would also contain off-diagonal elements. In 
\cite{Fejos2016,Fejos2017} it was suggested to first diagonalize the propagator matrix and only then regulate each eigenmode. This strategy led to the desired fixed point structure, but the main problem with such an approach is that since the rotation matrix that connects the two bases depend explicitly on the expectation value of the scalar field, it invalidates the Wetterich equation. Furthermore, if one transforms back the regulator matrix to the original basis, one finds that they break the $O(2N)$ global symmetry of the system, which may lead to the RG evolution of symmetry breaking terms in the effective action, making the whole procedure questionable.

Projecting (\ref{Eq:wet}) onto a homogeneous field configuration, one obtains the flow equation for the effective potential, $V_k(\rho)$:
\bea
&&k\partial_{k}V_{k}\left(\rho\right) =\frac{\Omega_{d}k^{d+2}}{d}\Bigg[\frac{\left(d-1\right)}{k^2+\frac{m_{A,k}^{2}}{Z_{A,k}}+2 Z_{\phi,k}e_k^{2}\rho}.\nonumber \\
 && \hspace{1.7cm}+\frac{2N-1}{k^{2}+\frac{V_k'(\rho)}{Z_{\phi,k}}(\rho)}+\frac{1}{k^{2}+\frac{V_k'(\rho)+2\rho V_k''}{Z_{\phi,k}}}\Bigg]\nonumber\\
 && \hspace{1.7cm}-\frac{\Omega_{d}k^{d+2}}{d\left(d+2\right)}\left[\frac{\left(d-1\right)}{k^{2}+\frac{m_{A,k}^{2}}{Z_{A,k}}+2 Z_{\phi,k} e_k^{2}\rho}\right]\eta_{A,k}\nonumber \\
 && -\frac{\Omega_{d}k^{d+2}}{d\left(d+2\right)}\left[\frac{2N-1}{k^{2}+\frac{V_k'(\rho)}{Z_{\phi,k}}(\rho)}+\frac{1}{k^{2}+\frac{V_k'(\rho)+2\rho V_k''}{Z_{\phi,k}}}\right]\eta_{\phi,k},\nonumber\\
\eea
where we defined the scalar and gauge anomalous dimensions as $\eta_{\phi,k}=-k\partial_k Z_{\phi,k}/Z_{\phi,k}$ and $\eta_{A,k}=-k\partial_k Z_{A,k}/Z_{A,k}$, respectively.

Fixed points are found in terms of dimensionless quantities. After introducing dimensionless variables,
\bea
\label{Eq:dimless}
\bar{V}_{k}&=&k^{-d}V_{k}, \quad \bar{\rho}=\rho k^{2-d}Z_{\phi,k} \nonumber\\
\bar{e}_{k}^{2}&=&e_{k}^{2}k^{d-4}, \quad \bar{m}^2_{A,k}=Z_{A,k}^{-1}k^{-2}m^2_{A,k},
\eea
the dimensionless flow equation for $\bar{V}_{k}\left(\bar{\rho}\right)$ is found to be
\bea
&&k\partial_{k}\bar{V}_{k}\left(\bar{\rho}\right) =-d\bar{V}_{k}\left(\bar{\rho}\right)+\left(d-2+\eta_{\phi,k}\right)\bar{\rho}\bar{V}_{k}'\left(\bar{\rho}\right)\nonumber \\
 && \hspace{1.6cm}+\frac{\Omega_{d}}{d}\Bigg[\frac{\left(d-1\right)}{1+\bar{m}_{A,k}^{2}+2\bar{e}_{k}^{2}\bar{\rho}}\nonumber\\
 &&\hspace{1.6cm}+\frac{2N-1}{1+\bar{V}_{k}'(\bar{\rho})}+\frac{1}{1+\bar{V}_{k}'(\bar{\rho})+2\bar{\rho}\bar{V}_{k}''(\bar{\rho})}\Bigg]\nonumber\\
 && \hspace{1.6cm}-\frac{\Omega_{d}}{d\left(d+2\right)}\left[\frac{\left(d-1\right)}{1+\bar{m}_{A,k}^{2}+2\bar{e}_{k}^{2}\bar{\rho}}\right]\eta_{A,k}\nonumber\\
 && -\frac{\Omega_{d}}{d\left(d+2\right)}\left[\frac{2N-1}{1+\bar{V}_{k}'\left(\bar{\rho}\right)}+\frac{1}{1+\bar{V}_{k}'\left(\bar{\rho}\right)+2\bar{\rho}\bar{V}_{k}''\left(\bar{\rho}\right)}\right]\eta_{\phi,k}.\nonumber\\
\eea
The flow of the scalar wave function renormalization factor, and therefore, the corresponding anomalous dimension is determined by the transverse projection of the two point function, $k\partial_k Z_{\phi,k}= \partial \Big[\frac{\delta^2 k\partial_k\Gamma_k}{\delta s_i \delta s_i}(p)\Big]/\partial p^{2}\big|_{p^2\rightarrow 0}$ ($i>1$). 

Applying the chain rule to (\ref{Eq:wet2}), and choosing $i=2$ for example, we get:
\bea
\label{Eq:gamma2flow}
&&\partial_k \frac{\delta^2 \Gamma_k}{\delta s_2(p)\delta s_2(-p)} = -\frac12 \int \tilde{\partial}_k\Bigg[ (\Gamma_k^{(2)}+{\cal R}_k)^{-1}_{A_iA_j} \nonumber\\
&&\frac{\delta \Gamma^{(2)}_{k,A_j\pi_2}}{\delta s_2(p)}(\Gamma_k^{(2)}+{\cal R}_k)^{-1}_{\pi_2 \pi_2} \frac{\delta \Gamma^{(2)}_{k,\pi_2A_i}}{\delta s_2(-p)}\Bigg]+\const \nonumber\\
&&+\sccon,
\eea
where we did not indicate explicitly the integrals over the Fourier variables. The constant comes from momentum independent tadpole diagrams, and it does not give contribution for the scalar wave function renormalization, while the scalar contributions are those arise from the ungauged $O(2N)$ model, see details in \cite{Delamotte2012}. The three point vertex appearing in (\ref{Eq:gamma2flow}), satisfying momentum conservation, is
\bea
\label{Eq:gamma3}
\Gamma^{(3)}_{A_i\pi_as_b}(q_1,q_2,-q_1-q_2)=i Z_{\phi,k}e(2q_2+q_1)_i\delta_{ab},
\eea
where $Z_{\phi,k}=Z_{e,k}$ is understood. Using (\ref{Eq:gamma3}) with the expressions for the regulated propagators (\ref{Eq:regprop}), after a somewhat long and tedious calculation, we get the following implicit equation for $\eta_{\phi,k}$:
\begin{widetext}
\bea
\eta_{\phi,k} =-8\bar{e}_{k}^{2}\varOmega_{d}\frac{\left(d-1\right)\left(4+2d-\eta_{\phi,k}-\eta_{A,k}+\left(2+d-\eta_{\phi,k}\right)\left(\bar{m}_{A,k}^{2}+2\bar{e}_{k}^{2}\bar{\rho}_0\right)\right)}{d^{2}(d+2)\left(1+\bar{m}_{A,k}^{2}+2\bar{e}_{k}^{2}\bar{\rho}_0\right)^{2}}\nonumber+\frac{4\varOmega_{d}}{d}\frac{\bar{\rho}_0\bar{V}_{k}''\left(\bar{\rho}_0\right)^{2}}{\left(1+2\bar{\rho}_0\bar{V}_k''\left(\bar{\rho}_0\right)\right)^{2}},\\
\eea
\end{widetext}
which is evaluated at the minimum point of the potential, $\bar{V}_{k}'\left(\bar{\rho}_0\right)=0$. Here, the second term is the usual scalar contribution, already present in the theory without the gauge field.

The flow of the gauge wave function renormalization is determined similarly, $k\partial_k Z_{A,k}= {\cal T}\partial \Big[\frac{\delta^2 k\partial_k\Gamma_k}{\delta A_i \delta A_j}(p)\Big]/\partial p^2\big|_{p^2\rightarrow 0}$,
where ${\cal T}$ projects out the coefficient of the $\delta_{ij} - \hat{p}_i\hat{p}_j$ transverse tensor structure\footnote{Here $\hat{p}_i=p_i/|p|$ is a unit vector.}. Note that, the longitudinal projection is not zero due to the presence of the regulator, but it is completely unimportant, as it can only induce a finite shift in the inverse of the gauge fixing parameter. If one adopts the Landau gauge, $\xi=0$, at the UV scale, i.e., $\xi^{-1}=\infty$, then no regular flow can modify this condition and thus $\xi_k^{-1}=\infty$ ($\xi_k=0$) remains true for any $k$. We will also come back to this point in \ref{sec:mWTI}, when analyzing the corresponding modified Ward-Takahashi identity.

Applying the chain rule again to (\ref{Eq:wet2}), we get
\bea
&&\partial_k\frac{\delta^2 \Gamma_k}{\delta A_i(p)\delta A_j(-p)}=-\int \tilde{\partial}_k\nonumber\\
&&\Bigg[(\Gamma_k^{(2)}+{\cal R}_k)^{-1}_{s_as_b}\frac{\delta \Gamma_{k,s_b\pi_b}^{(2)}}{\delta A_i(p)}(\Gamma_k^{(2)}+{\cal R}_k)^{-1}_{\pi_b\pi_a}\frac{\delta \Gamma_{k,\pi_as_a}^{(2)}}{\delta A_j(-p)}\nonumber\\
&&+(\Gamma_k^{(2)}+{\cal R}_k)^{-1}_{A_kA_l}\frac{\delta \Gamma_{k,A_ls_a}^{(2)}}{\delta A_i(p)}(\Gamma_k^{(2)}+{\cal R}_k)^{-1}_{s_as_b}\frac{\delta \Gamma_{k,s_bA_k}^{(2)}}{\delta A_j(-p)}\Bigg].\nonumber\\
&&+\const
\eea
Using (\ref{Eq:regprop}) and (\ref{Eq:gamma3}), and evaluating the integrals for small $p$, we arrive at
\bea
&&\eta_{A,k}=8\bar{e}_{k}^{2}\Omega_{d}\frac{N\left(1+2\bar{\rho}_0\bar{V}_{k}''\right)^{2}-4\bar{\rho}_0\bar{V}_{k}''\left(1+\bar{\rho}_0\bar{V}_{k}''\right)}{d\left(d+2\right)\left(1+2\bar{\rho}_0\bar{V}_{k}''\right)^{2}}\nonumber\\
&&\hspace{-0.5cm}+16\bar{e}_{k}^{4}\Omega_{d}\frac{\bar{\rho}_0\Big(d^{2}+\eta_{k}+\left(\eta_{k}-d\right)(\bar{m}_{A,k}^{2}+2\bar{e}_{k}^{2}\bar{\rho}_0)\Big)}{d^{2}\left(d+2\right)\left(1+2\bar{\rho}_0\bar{V}_{k}''\right)^{2}\left(1+\bar{m}_{A,k}^{2}+2\bar{e}_{k}^{2}\bar{\rho}_0\right)^{2}},\nonumber\\
\eea
where we, once again, work at the minimum point, $\bar{V}_k'(\bar{\rho}_0)=0$. Note that since we employ the constraint $Z_{\phi,k}=Z_{e,k}$, the flowing charge is simply $e_k=e/Z_{A,k}^{1/2}$, therefore, the anomalous dimension of the gauge field completely determines the flow of the charge. The dimensionless flow equation for $e^2$ simply becomes
\begin{equation}
k\partial_{k}\bar{e}_{k}^{2}=-(4-d)\bar{e}_{k}^{2}+\eta_{A,k}\bar{e}_{k}^{2},
\end{equation}
showing that in any nontrivial charged fixed point $\eta_{A}=4-d$ \cite{Herbut1996}.

At this point we note that the zero momentum limit ($p\to 0$) of the flow of the gauge two point function is nonvanishing for $k>0$, meaning that there is a finite gauge mass emerging and thus violating gauge symmetry. This is due to the fact that the FRG by construction includes a momentum cutoff, which is known to break gauge symmetry. According to the standard procedure, one starts with a UV action that contains a mass term for the gauge field, which gets eaten up by fluctuations as $k\rightarrow 0$. The scale dependence of such a mass term is dictated by the corresponding regulator modified Ward-Takahashi identity (mWTI), see the next section in detail.

In principle, one can also determine the flow of the gauge mass from the Wetterich equation, however, it turns out to be different from the result of the mWTI, except for a symmetric background \cite{Fejos2017}, where they do agree. Since the mWTI encodes the recovery of gauge symmetry at $k=0$, we decided to use the latter for the flow of the gauge mass. For details regarding the discrepancy between the flow equation and the mWTI, the reader is referred to Appendix B.

\section{Modified Ward-Takahashi identities}

The introduction of the infrared regulator $\mathcal{R}_k$ explicitly breaks the local $U(1)$ gauge invariance. To ensure that gauge symmetry is restored in the infrared limit ($k\to 0$), the scale dependent effective action $\Gamma_k[\Phi]$ must satisfy modified Ward-Takahashi identities \cite{Gies:2006wv}. Denoting the fluctuating superfield by $\hat{\Phi}=\{\hat{\phi}, \hat{A}_i\}$, the master equation for the mWTIs reads:
\bea
\label{Eq:mWTI_master}
&&\left\langle \delta_G S[\hat{\Phi}]\right\rangle -\int_{x}\left\langle \delta_G\hat{\Phi}\left(x\right)\right\rangle \frac{\delta\Gamma_k}{\delta\Phi(x)}\nonumber\\
&&\hspace{1.5cm}=-\delta_G\int_{x,y}\left\langle \hat{\Phi}\left(x\right){\cal R}_{k}\left(x,y\right)\hat{\Phi}\left(y\right)\right\rangle _{c},
\eea
where $\delta_G$ refers to an infinitesimal gauge transformation. We evaluate both sides of Eq.~\eqref{Eq:mWTI_master} in a uniform scalar background ($\partial_i \phi = 0$) up to linear order in the background gauge field $A_i$, which determines the photon mass as a function of $k$. In accordance with the previous section, we are working at the minimum point of the effective potential, $V_k'(\rho_0)=0$. Details of the calculation are found in Appendix A. We arrive at
\bea
\label{Eq:mA2}
m_{A,k}^{2}=-\Omega_{d}\frac{4 e^{2}k^{d}}{d(d+2)}\Bigg[&&\!\!\!\!\!\!\frac{1}{k^{2}+2\rho_0 V_{k}''\left(\rho_0\right)/Z_{\phi,k}}\nonumber\\
&&+\left(N-1\right)\frac{1}{k^{2}}\Bigg],
\eea
showing that for $d>2$, $m_{A,k}^2 \rightarrow 0$ as $k\rightarrow 0$, as it should. One can also read off from (\ref{Eq:mA2}) the UV value of the gauge mass at the initial scale by setting $k=\Lambda$. Using rescaled variables, (\ref{Eq:mA2}) becomes
\bea
 \bar{m}_{A,k}^{2}=-\Omega_{d}\frac{4\bar{e}_{k}^{2}}{d(d+2)}\left[\frac{1}{1+2\bar{\rho}_0\bar{V}_{k}''(\bar{\rho}_0)}+N-1\right],
\eea
which is valid at the minimum point of the rescaled potential $\bar{V}_{k}'\left(\bar{\rho}_0\right)=0$.

As for the $Z_{e}=Z_{\phi}$ identity, in the full theory it is followed from projecting (\ref{Eq:mWTI_master}) onto any quadratic combination of the scalar fields. Working in a non-zero background, i.e., $\langle s^{a}\rangle=v\delta_{a,1}$, however, we found that the truncation we are using leads to inconsistent momentum structure on the two sides of (\ref{Eq:mWTI_master}). When one works with the $\sim\! s^a s^b$ projection, one does arrive at the equality between the $Z$ factors at all scales ($Z_{e,k}=Z_{\phi,k}$), however, e.g., the $\sim \!s^a \pi^b$ projection fails to reproduce the same result. Since the $Z_{e}=Z_{\phi}$ is an exact relation in the full theory even in a non-zero background, the $Z_{e,k}=Z_{\phi,k}$ relation is adopted everywhere (as already announced in the previous section).

\section{Numerical fixed point solutions}

\begin{figure}
\includegraphics[scale=0.16]{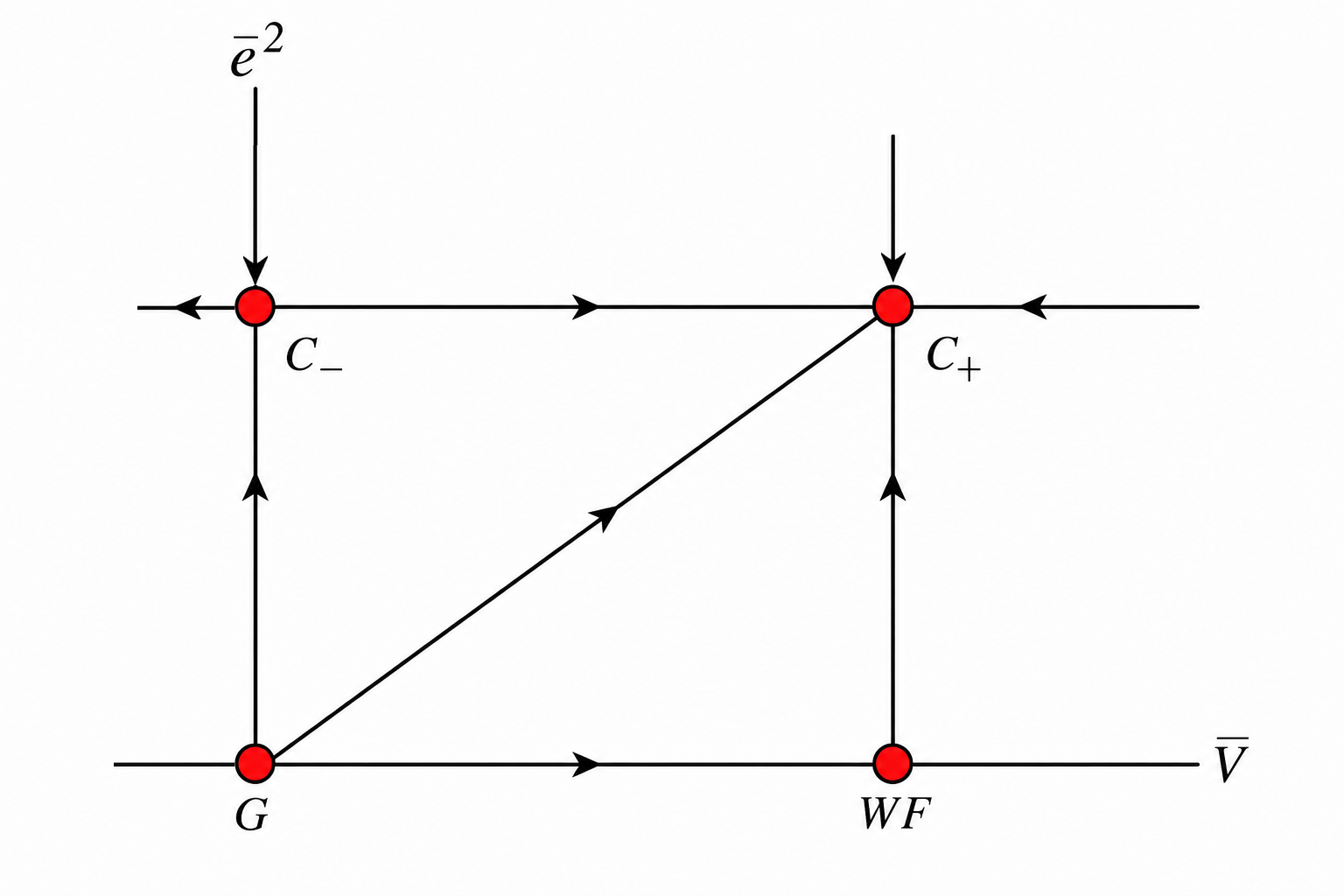}
\caption{Schematic of the RG flows (indicated by the arrows) and its fixed points (indicated by the red circles) deduced from the formal argument given below Eq. (\ref{eq:formal large-N FP eq2}) based on the conventional large $N$ scaling, corroborated near $d=4$ for $N\ge 183$ by $d=4-\epsilon$ expansion. G refers to the Gaussian, while WP to the Wilson-Fisher fixed point. For the definitions of $C_{+}$ and $C_{-}$, see the text. The temperature direction, relevant in all fixed points, can be considered perpendicular to the plane.} 
\label{fig:fpsketch}
\end{figure}

For finding fixed point potentials numerically, it is convenient to employ further rescalings. These also turn out to be helpful in obtaining large-N results. Introducing new variables (each carrying a tilde on top),
\bea
\bar{e}_{k}^{2}&=&d\Omega_d^{-1}\tilde{e}_{k}^{2}/N, \quad \bar{\rho}=\Omega_d N\tilde{\rho}/d, \nonumber\\
\bar{V}_{k}\left(\bar{\rho}\right)&=&\Omega_d N\tilde{V}_{k}\left(\tilde{\rho}\right)/d, \quad \bar{m}_{A,k}^{2}=\tilde{m}_{A,k}^{2},
\label{eq:scaling-Large-N}
\eea
the flow equation for the effective potential becomes
\bea
k\partial_{k}\tilde{V}_{k}\left(\tilde{\rho}\right) &=&-d\tilde{V}_{k}\left(\tilde{\rho}\right)+\left(d-2+\eta_{\phi,k}\right)\tilde{\rho}\tilde{V}_{k}'\left(\tilde{\rho}\right)\nonumber \\
 &+&\frac{\left(d-1\right)}{N\left(1+\tilde{m}_{A,k}^{2}+2\tilde{e}_{k}^{2}\tilde{\rho}\right)}+\frac{2-1/N}{1+\tilde{V}_{k}'\left(\tilde{\rho}\right)}\nonumber\\
 &+&\frac{1}{N\left(1+\tilde{V}_{k}'\left(\tilde{\rho}\right)+2\tilde{\rho}\tilde{V}_{k}''\left(\tilde{\rho}\right)\right)}\nonumber \\
 &-&\frac{\eta_{A,k}}{N\left(d+2\right)}\left[\frac{\left(d-1\right)}{1+\tilde{m}_{A,k}^{2}+2\tilde{e}_{k}^{2}\tilde{\rho}}\right]\nonumber \\
 &&\hspace{-2.2cm}-\frac{\eta_{\phi,k}}{\left(d+2\right)}\left[\frac{2-1/N}{1+\tilde{V}_{k}'\left(\tilde{\rho}\right)}+\frac{1}{N\left(1+\tilde{V}_{k}'\left(\tilde{\rho}\right)+2\tilde{\rho}\tilde{V}_{k}''\left(\tilde{\rho}\right)\right)}\right],\nonumber\\
 \label{eq:FPeq_tildeV}
\eea
while the anomalous dimensions are
\begin{widetext}
\bea
 \eta_{\phi,k} &\!\!\!=\!\!\!&-8\tilde{e}_{k}^{2}\frac{\left(d-1\right)\left(4+2d-\eta_{\phi,k}-\eta_{A,k}+\left(2+d-\eta_{\phi,k}\right)\left(\tilde{m}_{A,k}^{2}+2\tilde{e}_{k}^{2}\tilde{\rho}\right)\right)}{Nd^{2}(d+2)\left(1+\tilde{m}_{A,k}^{2}+2\tilde{e}_{k}^{2}\tilde{\rho}\right)^{2}}+4\frac{\tilde{\rho}_0\left(\tilde{V}_{k}''\left(\tilde{\rho}_0\right)\right)^{2}}{N\left(1+2\tilde{\rho}_0\tilde{V}_{k}''\left(\tilde{\rho}_0\right)\right)^{2}},\\
\eta_{A,k} &\!\!\!=\!\!\!&8\tilde{e}_{k}^{2}\frac{N\left(1+2\tilde{\rho}_0\tilde{V}_{k}''\left(\tilde{\rho}_0\right)\right)^{2}-4\tilde{\rho}_0\tilde{V}_{k}''\left(\tilde{\rho}_0\right)\left(1+\tilde{\rho}_0\tilde{V}_{k}''\left(\tilde{\rho}_0\right)\right)}{N\left(d+2\right)\left(1+2\tilde{\rho}_0\tilde{V}_{k}''\left(\tilde{\rho}_0\right)\right)^{2}}
+\frac{16\tilde{e}_{k}^{4}\tilde{\rho}_0\Big[d^{2}+\eta_{k}+\left(\eta_{k}-d\right)\left(\tilde{m}_{A,k}^{2}+2\tilde{e}_{k}^{2}\tilde{\rho}_0\right)\Big]}{Nd\left(d+2\right)\left(1+2\tilde{\rho}_0\tilde{V}_{k}''\left(\tilde{\rho}_0\right)\right)^{2}\left(1+\tilde{m}_{A,k}^{2}+2\tilde{e}_{k}^{2}\tilde{\rho}_0\right)^{2}}.\nonumber\\
\label{eq:eta_Ak_tilde}
\eea
\end{widetext}
Here $\tilde{\rho}_0$ is the minimum point of the $\tilde{V}_k$ potential, $\tilde{V}_k'(\tilde{\rho}_0)=0$. The flow of the charge is simply
\begin{equation}
k\partial_{k}\tilde{e}_{k}^{2}=-(4-d)\tilde{e}_{k}^{2}+\eta_{A,k}\tilde{e}_{k}^{2},
\label{eq:evl_tilde_ek2}
\end{equation}
while the mWTI for the gauge mass becomes
\begin{equation}
\tilde{m}_{A,k}^{2}=-\frac{4\tilde{e}_{k}^{2}}{(d+2)}\left[\frac{1}{N\left(1+2\tilde{\rho}_0\tilde{V}_{k}''\left(\tilde{\rho}_0\right)\right)}+\left(1-1/N\right)\right].
\label{eq:mA_tilde}
\end{equation}
Conventional large-$N$ results for the critical FP with one relevant direction can be found by formally setting $1/N \rightarrow 0$, 
\bea
k\partial_{k}\tilde{V}_{k}\left(\tilde{\rho}\right) &=&-d\tilde{V}_{k}\left(\tilde{\rho}\right)+\left(d-2+\eta_{\phi,k}\right)\tilde{\rho}\tilde{V}_{k}'\left(\tilde{\rho}\right)\nonumber \\
\label{eq:formal large-N FP eq1}
 &+&\Big(1-\frac{\eta_{\phi,k}}{d+2}\Big)\frac{2}{1+\tilde{V}_{k}'\left(\tilde{\rho}\right)},\\ 
\eta_{\phi,k}&=&0, \quad \eta_{A,k}=8\tilde{e}_k^2/(d+2),\\
\tilde{m}_{A,k}^{2}&=&-\frac{4\tilde{e}_{k}^{2}}{(d+2)}.
\label{eq:formal large-N FP eq2}
\eea
The flow equation for the charge remains unchanged, admitting a nontrivial FP solution, $\tilde{e}_*^2=(4-d)(d+2)/8$. Note that the flow of the scalar potential decouples from that of the charge, therefore, at $N=\infty$, it seems that for any non-charged fixed point there exists a charged counterpart, where the scalar potentials are the same, but the charge becomes finite. The counterpart of the Gaussian (G) fixed point is called $C_-$, while that of the Wilson-Fisher (WP) fixed point is $C_+$, see Fig. \ref{fig:fpsketch}.

One should be careful, since when taking the large-$N$ limit, the conventional scaling, given by Eq. (\ref{eq:scaling-Large-N}), was imposed, and it is known that, in the $O(N)$ and $O(N)\times O(2)$ models, this conventional scaling misses some of the multicritical FPs, where a boundary layer and thus a nonconventional scaling with respect to $1/N$ appear in the large-$N$ limit \cite{Yabunaka2017,Yabunaka2018,Yabunaka2022}. According to the aforementioned studies, the conventional scaling works for the usual critical fixed point with exactly one relevant direction. That said, if $d\lesssim 4$, then, indeed, two charged fixed points are found for $N\ge 183$ in the $d=4-\epsilon$ expansion, in addition to the G and WF fixed points, as shown in Fig. \ref{fig:fpsketch}. We will see that this structure remains in terms of global flows, for dimensions close but below $d=4$, however, when we deviate from $d=4$ towards $d=3$, it does not necessarily survive. The modified structure cannot be described by Eqs. (\ref{eq:formal large-N FP eq1})-(\ref{eq:formal large-N FP eq2}) under the conventional large-$N$ limit. The reason is that, as seen already in \cite{Fejos2017}, if one solves analytically the finite $N$ flow equations by approximating the scalar potential with its Taylor series around zero field, then the $C_-$ fixed point shows a different $N$-scaling compared the conventional one, and it turns out that its existence in terms of global potentials, at a given $N$, depends on the actual dimensionality.

At finite $N$, Eqs. \eqref{eq:FPeq_tildeV}--\eqref{eq:mA_tilde}  reduce  to the FP potential equation of the pure $O(2N)$ model without $U(1)$ gauge field, if $\tilde e^2_k=0$. Our strategy to find a charged FP is then as follows: We solve Eqs. \eqref{eq:FPeq_tildeV}--\eqref{eq:eta_Ak_tilde} and \eqref{eq:mA_tilde} by taking $\tilde e_k^2$ as an externally fixed parameter. We discretize the equations on the lattice space with $\tilde \phi\equiv \sqrt{2\tilde{\rho}}$ variable. We take the global $O(2N)$ Wilson-Fisher FP solution with $\tilde e_k^2=0$ as an initial condition and recursively solve Eqs. \eqref{eq:FPeq_tildeV}--\eqref{eq:eta_Ak_tilde} and \eqref{eq:mA_tilde} increasing $\tilde e_k^2$ slowly until Eq. \eqref{eq:evl_tilde_ek2} is satisfied. At each step, we use the solution of the previous step as an initial condition. We successfully performed this procedure for $N=125$ and $d=3.8$ and we found a charged FP solution of Eqs. \eqref{eq:FPeq_tildeV}--\eqref{eq:mA_tilde} with one relevant direction, which should describe a second-order phase transition, controlled by the temperature parameter. This is the $C_+$ FP, defined earlier already. Starting from this $C_+$ solution, we recursively solve  Eqs. \eqref{eq:FPeq_tildeV}--\eqref{eq:mA_tilde} varying $N$ and $d$ to explore the FP structure in the $(d,N)$-space. 

\begin{figure}
\includegraphics[scale=0.5]{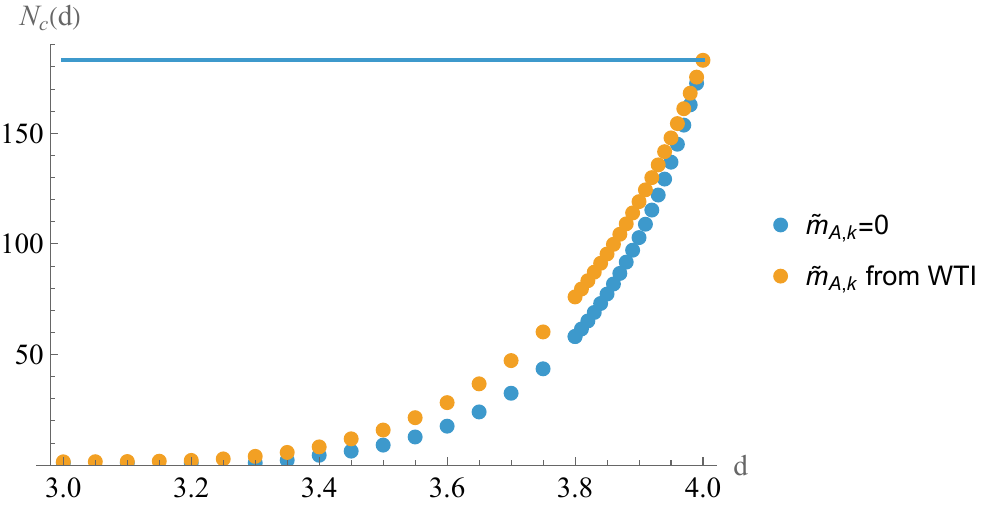}

\includegraphics[scale=0.5]{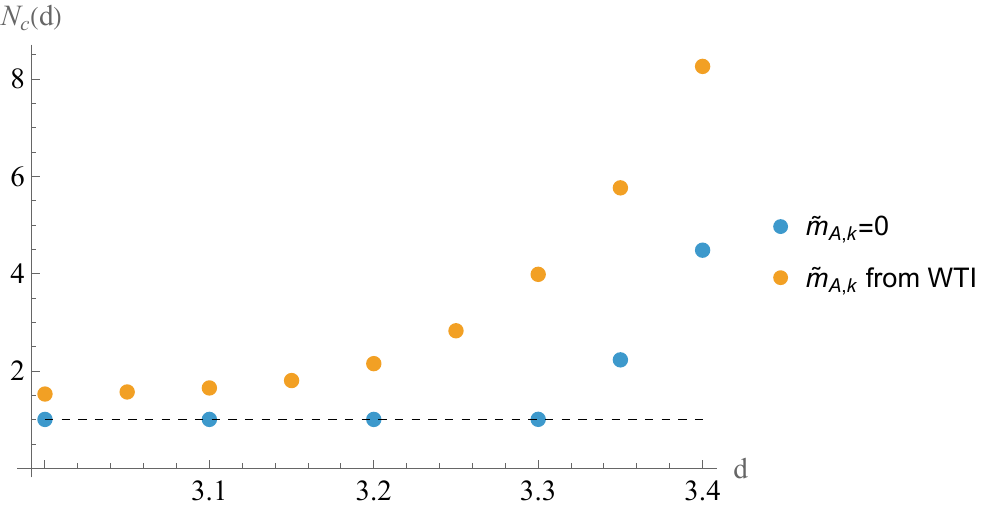}
\caption{The critical flavor number $N_c(d)$ as a function of spatial dimension $d$. (Top) $N_c(d)$ for $3\le d \le 4$. The horizontal line indicates $N_c\simeq183$, obtained at the leading order of the $\varepsilon$ expansion. (Bottom) $N_c(d)$ for $3\le d \le 3.4$. The dashed line indicates $N=1$.} 
\label{fig:Nc}
\end{figure}

The $C_+$ FP is found to exist for $N\ge N_c(d)$, whose curve is plotted in Fig. \ref{fig:Nc}, for each spatial dimension $d$. The standard result at the leading order of the $\epsilon=4-d$ expansion is $N_c(d) \simeq 183$. We found that our numerical value for $N_c(d)$, arising from a non-perturbative treatment, approaches this value as $d\to 4$. However, it is important to stress that $N_c(d)$ decreases rapidly as $d$ decreases. This shows that the reliability of the $\epsilon$ expansion rapidly fails when one deviates the dimensionality from $d=4$. In particular, at $d=3$ it gives a completely false result. We find that the critical flavor number at which $C_+$ disappears is $N_c(d=3) \simeq 1.5$, which is still above $1$, even though it is known that there does exist a second-order transition for strongly coupled conventional superconductors\footnote{The corresponding critical value for the Ginzburg-Landau parameter, that is, the ratio between the London penetration depth and the coherence length, is $\kappa\simeq 0.76/\sqrt2$ \cite{Mo2002}.}, i.e., one expects $N_c(d=3)<1$. We suspect that this value could be improved by extending our truncation, but we wish to highlight that it is already very close to 1, much closer than the predictions of earlier studies, showing that a functional treatment is indeed necessary to properly mapping the structure of fixed points. It would be interesting to perform a complete leading order calculation of the effective action in terms of the derivative expansion, but we leave that to future work.

\begin{figure}
\includegraphics[scale=0.5]{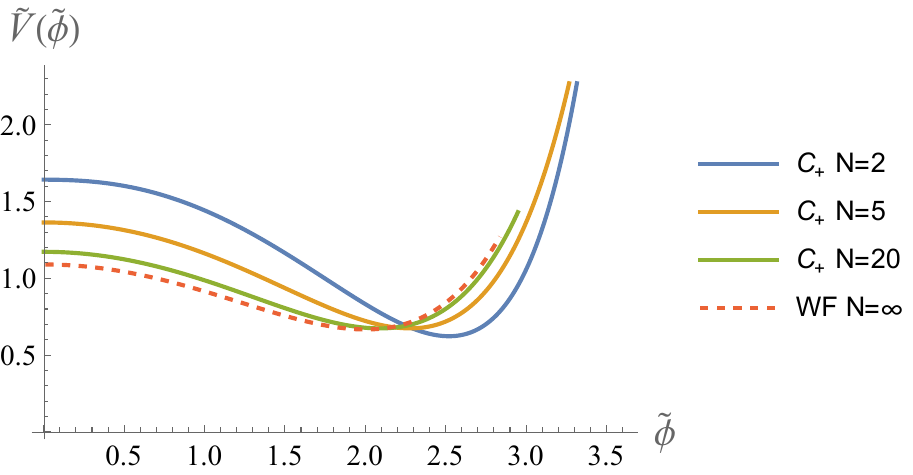}

\caption{Global fixed point potential $\tilde{V}(\tilde{\phi})$ for the $C_+$ FP in $d=3$. In the $N\to \infty$ limit, it approaches the FP potential for WF FP.} 
\label{fig:FPpot}
\end{figure}

For $N \gtrsim N_c(d)$, perturbing the solution for $C_+$ and using it as an initial condition, another charged FP can be found, with two relevant directions, which is the $C_-$ FP. At $N=N_c(d)$, $C_+$ and $C_-$ collide and disappear. This is consistent with Fig. 1 of \cite{Fejos2017}, except that all fixed points are obtained globally. We confirmed that, in the $N\to \infty$ limit, the scalar potential at the $C_+$ FP approaches that of the WF FP, as shown in Fig. \ref{fig:FPpot}. For small $\epsilon=4-d>0$ and given that $N>N_c(d)$, we also confirmed that there are four global FPs, i.e., $C_+$, $C_{-}$, $O(2N)$ Wilson-Fisher, and the Gaussian FP, which is consistent with the leading order results of the $\epsilon$ expansion.

When working in $d=3$, however, the structure described above gets modified, signaling that the $\epsilon$ expansion cannot capture all details of the fixed points in the system. While the structure seen in Fig. \ref{fig:fpsketch} remains for $1.5\lesssim N\lesssim 9.5$, we find that when increasing $N$ from $N_c(d=3)\simeq 1.5$, around $N\simeq9.5$, $C_-$ collides with another, yet unknown FP with three unstable directions and disappears. This nontrivial behavior of the bicritical $C_-$ FP is quite similar to that found at the large-$N$ limit of $O(N)\times O(2)$ models, where a perturbative multicritical FP disappears at large-$N$ by colliding another nonperturbative multicrical FP, which cannot be found perturbatively. Since $C_-$ exists at large-$N$ for $d\lesssim 4$, one can also check that starting from some large enough $N$ and decreasing the dimensionality from $d\approx 4$ towards $d\rightarrow 3$, $C_-$ indeed disappears close to $d=3$.

Fixed point collisions and the structure of multicritical fixed points in the Abelian Higgs model are certainly very interesting, but since our focus in this study is to confirm the existence of $C_+$ and thus a second-order phase transition, we leave a more complete study of the large-$N$ structure of multicritical FPs in Abelian Higgs models to future work.

\begin{figure}
\includegraphics[scale=0.54]{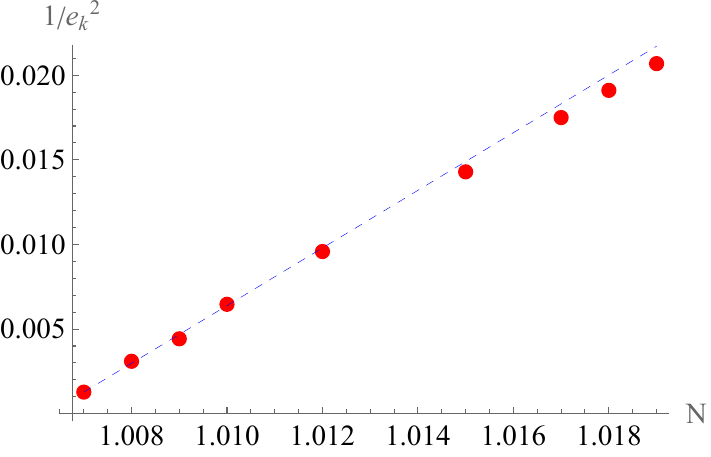}
\caption{The inverse of the scaled squared charge, $1/\tilde{e}_k^2$, at the $C_+$ FP as a function of the flavor number $N$ in $d=3$ when setting $\tilde{m}_{A,k}$=0. The dashed line refers to the line $\tilde{e}_k^2=0.587/(N-1.00625)$.} 
\label{fig:ek}
\end{figure}

Finally, we also want to draw attention on a different approach with regards to the gauge mass. Note that, while the dimensionful gauge mass disappears in the $k\to 0$ limit, the dimensionless one is nonvanishing at any charged fixed point. One may argue that even though it is a direct consequence of the mWTI, a nonvanishing dimensionless gauge mass is only a reminiscent artifact of the cutoff regularization, and should not be taken physical. Therefore, on top of solving the coupled Eqs. \eqref{eq:FPeq_tildeV}--\eqref{eq:mA_tilde}, we also worked out the scenario when the dimensionless gauge mass is set to zero by hand, $\tilde{m}_{A,k}^2=0$. The corresponding $N_c(d)$ curve can also be seen in Fig. \ref{fig:Nc}. Below $d\simeq 3.3$, $C_+$ continues to exist without colliding with $C_{-}$ very close to $N=1$, but for $d=3$, the dimensionless charge $\tilde{e}_k$ seems to diverge at $N\simeq 1.00625$, as shown in Fig. \ref{fig:ek}. Presumably this is, once again, an artifact of the LPA' approximation, and it would be interesting to see whether the charge can remain regular in a more sophisticated truncation.

\section{Summary}

In this paper, we investigated the fixed point structure of the Abelian Higgs model with $N$ complex scalars, via the search for global potentials, using the functional renormalization group approach. Our main motivation was to see whether the FRG can accommodate contradictions between the perturbative RG and Monte Carlo simulations. According to the latter, for $N=1$, the superconducting phase transition turns from first to second order as the self-coupling of the scalar field increases. Simulations also showed that for $2\leq N \lesssim 10$, the transition seems to always be first order, while for larger flavor numbers the second order nature reappears for strong self-couplings. The main contradiction between these findings and those of the leading order perturbative RG results arises from the fact that the $\beta$ functions show no signs of an infrared-stable fixed point or its bicritical counterpart, both of which are needed for the existence of a tricritical point, unless $N>183$. We note here that the actual transition order is also subject to the domain of attraction of a critical fixed point, therefore, the mere existence of the latter does not guarantee critical behavior.

After deriving flow equations for the non-perturbative scalar potential and the charge in the LPA', we searched for global fixed points for various $d$ and $N$ values. First, we confirmed that for $d\approx 4$, the non-perturbative approach is consistent with its perturbative counterpart, as expected. However, on the contrary to the usual $O(N)$ model, for the Abelian Higgs model we found that as we deviate from $d=4$ toward $d=3$, the non-perturbative results start to differ significantly.

In $d=3$, we found that for $N>N_c(d=3)\simeq 1.5$, there is a critical fixed point with nonzero charge ($C_+$), having one relevant direction, together with another charged fixed point $(C_-)$ with two relevant directions. Interestingly, while this fixed point collides with $C_+$ at $N=N_c(d=3)$, it can also collide with a yet unknown fixed point with three relevant directions at $N\simeq 9.5$. We showed that this scenario is very specific to $d=3$ and invalidates the conventional large-$N$ scaling. We also found that when the dimensionality is increased, both $C_+$ and $C_-$ exist for any $N>N_c(d)$, and thus the validity of the conventional large-$N$ scaling is restored.

The fact that below $N=N_c(d=3)\simeq 1.5$ we find no charged fixed points shows that our approach is not capable of giving account of the tricritical point for $N=1$, since the two charged fixed points necessary exist only the interval $1.5\lesssim N \lesssim 9.5$. In this interval we have both first and second order transitions depending on the Ginzburg-Landau parameter, which characterizes the relative strength between the gauge-scalar and scalar-scalar interactions. As $C_-$ disappears at $N\simeq 9.5$, the critical $C_+$ fixed point remains, and it is an open question what its domain of attraction becomes and therefore the regions of the parameter space belonging to a second order transition. In light of other potentially multicritical fixed points that might exist in the system, this is highly nontrivial. It would certainly be important to find more fixed points and to check how the size of the above interval and the value for $N_c(d=3)$ vary when the truncation is improved. One is especially interested in comparing $N_c(d=3)$ with the predictions of Monte Carlo simulations.

We also investigated the case, where the gauge mass, whose flow is dictated in principle by the modified Ward-Takahashi identity and is not vanishing at the charged fixed points, is taken to be zero by hand. This was motivated by the fact that even though the mWTI forces us to introduce a gauge mass, one might argue that it is only a remnant of the cutoff regularization and should not be considered physical at any fixed point. What we found is that as we approach $d\rightarrow 3$, indeed $N_c(d)\rightarrow 1$ in $d=3$. Unfortunately, the exact limit could not be taken, as the charge seems to diverge in the $C_+$ FP slightly above $N=1$.

As mentioned above, it would be important to improve the truncation to see the robustness of our results. A natural extension would be to perform a complete leading order calculation in terms of the derivative expansion. This is significantly more complicated than the LPA', but it was successfully worked out in the past for $O(N)$ scalar models, and we expect that it is also feasible for the Abelian-Higgs model. Furthermore, since we identified a new type of fixed point collision for $C_-$, it would also be interesting to investigate whether there exist more multicritical FPs in the system, similarly to $O(N)$ and $O(N)\times O(2)$ models \cite{Yabunaka2017,Yabunaka2018,Yabunaka2022}. In particular, since the aforementioned collision leads to the invalidity of the conventional large-$N$ limit in $d=3$, exploration in this direction would also be important in understanding under what conditions a $1/N$ expansion is sensible. These directions represent future work, to be reported elsewhere.

\section*{Acknowledgments}

We are grateful for Igor Herbut and Adam Nahum for their useful remarks concerning the manuscript. G.F. was supported by the Hungarian National Research, Development, and Innovation Fund under Project No. FK142594 and by the Fulbright Program, sponsored by the U.S. Department of State. S.Y. was supported by JSPS Grant-in-Aid for Scientific Research (Grants Nos. 21K03488 and 24K06984).

\appendix
\renewcommand{\theequation}{A\arabic{equation}}
\setcounter{equation}{0}

\section{Evaluation of the modified Ward-Takahashi Identity for the gauge mass}
\label{sec:mWTI}

Here we evaluate the projection of Eq. (\ref{Eq:mWTI_master}) onto linear terms in the gauge field, which leads to the mWTI for the gauge mass. Since the bare action $S[\hat{\Phi}]$ is gauge invariant except for the gauge fixing term, the classical variation yields
\bea
\langle \delta S[\hat{\Phi}] \rangle = Z_{A,\Lambda}\xi^{-1} \int_{x} \partial_{j}A_{j}(x)\partial_i\partial_i \theta(x)
\eea
where we used that upon gauge transformation, $\delta A_i(x) = - \partial_i \theta(x)$. To evaluate the variation of the effective action, we calculate the gauge variation of each component in $\Gamma_k$ individually. For the scalar kinetic term, taking into account that $\rho$ is gauge invariant, i.e., $\delta \rho = 0$, we find:
\bea
\delta \int_{x} Z_{\phi,k}\partial_{i}\phi^{\dagger}\partial_{i}\phi &=& -ie \int_{x} \theta \Big\{ Z_{\phi,k} \left( \partial_i\partial_i\phi^{\dagger}\phi - \phi^{\dagger}\partial_i\partial_i\phi \right) \notag \\
&+& Z_{\phi,k}' \partial_i \rho \left( \partial_{i}\phi^{\dagger}\phi - \phi^{\dagger}\partial_{i}\phi \right) \Big\},
\eea
where we used that upon gauge transformation, $\delta \phi(x)=ie\theta(x)\phi(x)$, and the prime denotes the derivative with respect to $\rho$. The variation of the gauge term in $\Gamma_k$ yields:
\begin{align}
&\delta \int_{x} \frac{Z_{A,k}}{2} A_{i} \left[ -\partial^{2}\delta_{ij} + \left(1-\xi_k^{-1}\right)\partial_{i}\partial_{j} \right] A_{j} \notag \\
&= Z_{A,k}\xi_k^{-1} \int_{x}\partial_{j}A_{j} \partial_i\partial_i\theta.
\end{align}

For the three-point interaction term, variations with respect to both the gauge field and the scalar fields must be collected:
\bea
&&\hspace{-0.8cm}\delta \int_{x} \Big[-ie Z_{e,k} \left(\partial_{i}\phi^{\dagger}\phi - \partial_{i}\phi\phi^{\dagger}\right)A_{i}\Big] \notag \\
&&\hspace{-0.8cm}= \int_{x} \theta \Big\{ -ie Z_{e,k} \partial_{i} \left(\partial_{i}\phi^{\dagger}\phi - \partial_{i}\phi\phi^{\dagger}\right) \notag \\
&&\hspace{-0.3cm} - Z_{e,k}' \partial_{i}\rho \left(\partial_{i}\phi^{\dagger}\phi - \partial_{i}\phi\phi^{\dagger}\right)+2e^{2} \partial_{i} \left( Z_{e,k}\rho A_{i} \right) \Big\}. 
\eea
Finally, the variation of the four-point interaction and the photon mass term gives:
\begin{align}
\delta \int_{x} \frac{Z_{e,k}^{2}}{Z_{\phi,k}}e^{2}A_{i}^{2}\phi^{\dagger}\phi 
&= -\int_{x} \theta \partial_i \left( 2\frac{Z_{e,k}^{2}}{Z_{\phi,k}}e^{2}\rho A_{i} \right), \\
\delta \int_{x} \frac{1}{2}m_{A,k}^{2}A_{i}^{2} 
&= -\int_{x} \theta \partial_i \left( m_{A,k}^{2} A_{i} \right).
\end{align}

Collecting all contributions for a uniform background ($\partial_i \rho = 0$) and keeping terms up to linear order in $A_i$, the left-hand side (LHS) of the mWTI simplifies to:
\begin{align}
(\text{LHS}) = \int_{x} \theta \Big\{ & \partial_{i}\left[ \left(m_{A,k}^{2} + 2Z_{e,k}\left(\frac{Z_{e,k}}{Z_{\phi,k}}-1\right)e^{2}\rho\right)A_{i} \right] \notag \\
& + \xi^{-1}Z_{A,\Lambda}\partial_{j}A_{j}\partial_i\partial_i \theta \notag \\
& - \xi_{k}^{-1}Z_{A,k}\partial_{j}A_{j}  \partial_i\partial_i\theta \Big\}.
\end{align}
Passing to momentum space ($\partial_i \to ip_i$, $\partial_i\partial_i \to -p^2$), we obtain the final form for the LHS:
\bea
\label{eq:LHS_final}
(\text{LHS}) &=& \int_{p} \theta(-p)A_{i}(p)\nonumber\\
&&\times\Big\{ ip_{i}\left(m_{A,k}^{2} + 2Z_{e,k}\left(\frac{Z_{e,k}}{Z_{\phi,k}}-1\right)e^{2}\rho\right) \nonumber\\
&&\hspace{0.3cm}- ip^{2}p_{i}\left(\xi^{-1} Z_{A,\Lambda} - \xi_{k}^{-1}Z_{A,k}\right) \Big\}.
\eea

The right-hand side (RHS) of the master equation of the mWTIs, i.e. Eq. (\ref{Eq:mWTI_master}), arises from the gauge variation of the scalar regulator terms. Decomposing the complex scalar field into real and imaginary parts, $\phi^a = (s^a + i\pi^a)/\sqrt{2}$, the appropriate gauge transformations become $\delta s^a(x) = -e\theta(x)\pi^a(x)$ and $\delta\pi^a(x) = e\theta(x) s^a(x)$. The trace over the regulator insertion is written as:
\begin{align}
(\text{RHS}) &= -\frac{1}{2} \sum_{a=1}^{N} \int_{q} Z_{\phi,k} R_{k}(q) \notag \\
&\quad \times \delta \langle \hat{s}^{a}(q)\hat{s}^{a}(-q) + \hat{\pi}^{a}(q)\hat{\pi}^{a}(-q) \rangle_{c},
\end{align}
where $R_k(q)$ is chosen to be the Litim regulator, see (\ref{Eq:reg}). Performing the gauge variation of the fields and changing the dummy integration variables, we can isolate the gauge parameter in momentum space as $\theta(-p)$:
\begin{align}
(\text{RHS}) &= e \sum_{a=1}^{N} \int_{p,q} \theta(-p) Z_{\phi,k}R_{k}(q) \notag \\
&\quad \times \left[ \langle \hat{\pi}^{a}(-p-q)\hat{s}^{a}(q) \rangle_{c} - \langle \hat{s}^{a}(-p-q)\hat{\pi}^{a}(q) \rangle_{c} \right].
\end{align}
Note that the connected two-point functions inside the bracket, $\langle \hat{\pi}^{a}(-p-q)\hat{s}^{a}(q) \rangle_{c}$ and $\langle \hat{s}^{a}(-p-q)\hat{\pi}^{a}(q) \rangle_{c}$, have a non-vanishing total momentum (the sum of the two momentum arguments is $-p \neq 0$) due to the presence of the external gauge field $A^i(p)$. 

To evaluate these non-diagonal correlation functions with explicit momentum dependencies, we expand them up to linear order in $A^i(p)$. Utilizing the full propagators and the three-point vertex function where all momentum indices are strictly maintained, we decompose them into the diagonal (vanishing total momentum) propagators $G_{\pi^a\pi^a}(q)$ and $G_{s^as^a}(q)$ as follows:
\bea
\!\!\!\!\!\!\!\langle \hat{\pi}^{a}(-p-q)\hat{s}^{a}(q) \rangle_{c} &=& - G_{\pi^a\pi^a}(-p-q) \nonumber\\
&&\hspace{-2cm}\times \Gamma^{(3)}_{\pi^a A_i s^a}(-p-q, p, q) A^i(p) G_{s^as^a}(q), \\
\!\!\!\!\!\!\!\langle \hat{s}^{a}(-p-q)\hat{\pi}^{a}(q) \rangle_{c} &=& - G_{s^as^a}(-p-q) \nonumber\\
&&\hspace{-2cm}\times \Gamma^{(3)}_{s^a A_i \pi^a}(-p-q, p, q) A^i(p) G_{\pi^a\pi^a}(q).
\eea
Here, the three-point vertices with all incoming momenta are coming from (\ref{Eq:gamma3}):
\begin{align}
\Gamma^{(3)}_{\pi^a A_i s^a}(-p-q, p, q) &= -ie Z_{e,k} (2q+p)_i, \\
\Gamma^{(3)}_{s^a A_i \pi^a}(-p-q, p, q) &= ie Z_{e,k} (2q+p)_i.
\end{align}
Substituting these vertex functions back and using the momentum inversion symmetry of the diagonal propagators, $G(-p-q) = G(p+q)$, we find the explicit pre-integration expression for the RHS:
\bea
(\text{RHS}) &\!\!\!=\!\!\!& -i e^2 Z_{e,k} \sum_{a=1}^{N} \int_{p,q} \theta(-p) A^i(p) (2q+p)_i Z_{\phi,k}R_{k}(q) \notag \\
&&\hspace{-1.5cm}\times \Big[ G_{\pi^a\pi^a}(p+q) G_{s^as^a}(q)+ G_{s^as^a}(p+q) G_{\pi^a\pi^a}(q) \Big].
\eea
For a uniform background, where the symmetry is broken in the $a=1$ direction, the propagators are given by:
\bea
&&\hspace{-0.8cm}G_{s^a s^b}(q) =\left[ Z_{\phi,k} q^2 + R_k(q) + \delta_{a,1}2\rho_0 V_k''(\rho_0) \right]^{-1}\delta_{ab}, \nonumber\\ \\
&&\hspace{-0.8cm}G_{\pi^a \pi^b}(q)= \left[ Z_{\phi,k} q^2 + R_k(q) \right]^{-1}\delta_{ab},
\eea
where we worked at the minimum point of the potential, $V_k'(\rho_0)=0$. Employing the Litim regulator and expanding the integrand in powers of the external momentum $p$, by performing the loop integration over the internal momentum $q$, the RHS reduces to:
\begin{align}
\label{eq:RHS_final}
(\text{RHS}) &= i\Omega_{d}\frac{4Z_{e,k}e^{2}k^{d-2}}{Z_{\phi,k}d(d+2)} \int_{p} \theta(-p)A^{i}(p)p_{i} \notag \\
&\quad \times \Bigg( -k^{2}\bigg[ \frac{1}{k^{2}+2\rho_0 V_{k}''(\rho_0)/Z_{\phi,k}} + \frac{N-1}{k^{2}} \bigg] \notag \\
&\quad + p^{2}\bigg[ \frac{k^{2}+\rho_0 V_{k}''(\rho_0)/Z_{\phi,k}}{\left(k^{2}+2\rho_0 V_{k}''(\rho_0)/Z_{\phi,k}\right)^{2}} + \frac{N-1}{k^{2}} \bigg] \Bigg).
\end{align}
For the mWTIs to hold, the coefficients of the distinct momentum structures $p_i$ and $p^2 p_i$ must match identically between Eq.~\eqref{eq:LHS_final} and Eq.~\eqref{eq:RHS_final}. 

%


Comparing the terms proportional to $p_i$, we extract the constraint on the scale dependent photon mass parameter:
\bea
m_{A,k}^{2} &\!\!\!+\!\!\!& 2Z_{e,k}\left(\frac{Z_{e,k}}{Z_{\phi,k}}-1\right)e^{2}\rho_0 =-\Omega_{d}\frac{4Z_{e,k}e^{2}k^{d}}{Z_{\phi,k}d(d+2)} \nonumber\\
&&\hspace{0.4cm}\times \left[ \frac{1}{k^{2}+2\rho_0 V_{k}''(\rho_0)/Z_{\phi,k}} + \frac{N-1}{k^{2}} \right],
\eea
which leads to Eq. (\ref{Eq:mA2}) after setting $Z_{e,k}=Z_{\phi,k}$ at all scales.

Similarly, comparing the $p^2 p_i$ components yields the constraint on the gauge-fixing parameter and the gauge wave function renormalization:
\bea
\label{Eq:Z_xi_con}
-\left(\xi^{-1} Z_{A,\Lambda} - \xi_{k}^{-1}Z_{A,k}\right)&=& \Omega_{d}\frac{4Z_{e,k}e^{2}k^{d-2}}{Z_{\phi,k}d(d+2)} \nonumber\\
&&\hspace{-3.6cm} \times \left[ \frac{k^{2}+\rho_0 V_{k}''(\rho_0)/Z_{\phi,k}}{\left(k^{2}+2\rho_0 V_{k}''(\rho_0)/Z_{\phi,k}\right)^{2}} + \frac{N-1}{k^{2}} \right].
\eea
Eq. (\ref{Eq:Z_xi_con}) shows that the Landau gauge is indeed a fixed point of the RG flows; if we set $\xi=0$ at the UV scale, then $\xi_k=0$ must remain throughout the flow. We also note in passing that the overall sign on the right-hand side of the corresponding wave function identity reported in Ref.~\cite{Fejos2017} should be inverted to match this derivation.

\renewcommand{\theequation}{B\arabic{equation}}
\setcounter{equation}{0}

\section{Derivation of the flow equation for the gauge mass}
\label{app:photon_mass_flow}

In this appendix, we provide detailed derivation of the flow for the scale-dependent gauge field mass parameter, arising from the Wetterich equation. Evaluating the second functional derivative of (\ref{Eq:wet}) with respect to the gauge field $A_i$ at zero external momentum, we arrive at
\bea
\label{Eq:massflow}
k\partial_{k}\left(m_{A,k}^{2}+\frac{Z_{e,k}^{2}}{Z_{\phi,k}}e^{2}\rho\right) \delta^{ij}&=& k\tilde{\partial}_{k} \frac{\delta^{2}\varGamma_{k}}{\delta A_{i}(p)\delta A_{j}(-p)}\Bigg|_{p \to 0}.\nonumber\\
\eea

The right-hand side systematically decomposes into three topologically distinct diagrammatic contributions: the bubble diagrams mediated by the $A\mathchar`-s\mathchar`-\pi$ vertices, the bubble diagrams mediated by the $A\mathchar`-A\mathchar`-s$ vertices (proportional to the background $v^2$), and the tadpole diagrams originating from the $A\mathchar`-A\mathchar`-s\mathchar`-s$  or $A\mathchar`-A\mathchar`-\pi\mathchar`-\pi$ vertices. 

Since $k\tilde{\partial}_k$ acts exclusively on the explicit regulator dependence in the full propagators $G = (\Gamma_k^{(2)} + {\cal R}_k)^{-1}$, we use the chain rule identity $\tilde{\partial}_k G = -G \partial_k {\cal R}_k G$. Writing out the trace over all $2N$ internal scalar degrees of freedom and the gauge fields, the right hand side of (\ref{Eq:massflow}) yields
\bea
    \text{(RHS)} &=& 4e^{2}Z_{e,k}^{2}\int_{q} q^{2} k\partial_k (Z_{\phi,k}R_k(q)) \notag \\
    &&\hspace{-1.7cm} \times \Big( G_{s^1s^1}^2(q) G_{\pi\pi}(q) + G_{s^1s^1}(q) G_{\pi\pi}^2(q) + 2(N-1) G_{\pi\pi}^3(q) \Big) \nonumber\\
    &&\hspace{-1.1cm}+ 4 \frac{Z_{e,k}^{4}e^{4}v^{2}}{Z_{\phi,k}^{2}} \frac{d-1}{d} \int_{q}\Big[k\partial_k (Z_{A,k} R_k(q))G_A^2(q) G_{s^1s^1}(q) \nonumber\\
    &&\hspace{1.9cm}+ k\partial_k (Z_{\phi,k} R_k(q)) G_A(q) G_{s^1s^1}^2(q)\Big] \notag \\
    &&\hspace{-1.7cm}- e^{2} \frac{Z_{e,k}^{2}}{Z_{\phi,k}} \int_{q} k\partial_k (Z_{\phi,k} R_k(q))\Big( G_{s^1s^1}^2(q) + (2N-1) G_{\pi\pi}^2(q) \Big).\nonumber\\
\eea
Here $G_A(q)$ is the gauge propagator in momentum space, without the transverse tensor structure, $G_A(q)=(Z_{A,k}q_R^2+m_{A,k}^2+2Z_{e,k}^2e^2\rho/Z_{\phi,k})^{-1}$, and we used that $G_{s^as^a}=G_{\pi^i\pi^i}=G_{\pi\pi}$ $(a>1)$. Note that the inherent minus signs from the bubble loop traces exactly cancel against the chain rule derivative of the propagator, yielding strictly positive bubble contributions, whereas the tadpole loops become negative. The overall factor $(d-1)$ emerges from the transverse trace of the internal gauge propagator.

Using that $R_k(q)=(k^2-q^2)\Theta(k^2-q^2)$, we can analytically evaluate the two master radial integrals:
\begin{subequations}
\bea
    &&\hspace{-1cm}\int_{q} q^{2} k\partial_k(Z_{\phi/A,k}R_k(q)) \nonumber\\
    &=&\int_q [ 2k^{2} - \eta_{\phi/A,k}(k^{2}-q^{2})]\Theta(k^2-q^2)\nonumber\\
    &=& 2\varOmega_{d} k^{d+4} \frac{(d+4-\eta_{\phi/A,k})}{(d+2)(d+4)}, \\
    &&\hspace{-1cm}\int_{q} k\partial_k(Z_{\phi/A,k}R_k(q))\nonumber\\
    &=&\int_q  [ 2k^{2} - \eta_{\phi/A,k}(k^{2}-q^{2}) ]\Theta(k^2-q^2)\nonumber\\
    &=& 2\varOmega_{d} k^{d+2} \frac{(d+2-\eta_{\phi/A,k})}{d(d+2)}.
\eea
\end{subequations}

Assuming that at all scales $Z_{\phi,k}=Z_{e,k}$, we arrive at the following flow equation:
\begin{widetext}
\bea
\label{Eq:massflow_final}
k\partial_{k} \left( {m}_{A,k}^{2} + 2Z_{\phi,k} {e}_k^2 {\rho}_0) \right)&\!\!\!=\!\!\!&8\Omega_dk^{d+4} {e}_k^2 \frac{d+4-\eta_k}{d(d+2)(d+4)}\Bigg[\frac{2(N-1)}{k^6}+\frac{1}{k^2\left(k^2+\frac{2{\rho}_0 {V}_k''({\rho}_0)}{Z_{\phi,k}}\right)^2}+ \frac{1}{k^4\left(k^2+\frac{2{\rho}_0{V}_k''({\rho}_0)}{Z_{\phi,k}}\right)}\Bigg]\nonumber\\
&&\hspace{-3.5cm}+16\Omega_dk^d {e}_k^4 \frac{d-1}{d^2(d+2)}{\rho}_0 \Bigg[\frac{d+2-\eta_k}{k^2+\frac{{m}_{A,k}^2}{Z_{A,k}}+2Z_{\phi,k}{e}_k^2{\rho}_0}\frac{1}{\left(k^2+\frac{2{\rho}_0 {V}_k''({\rho}_0)}{Z_{\phi,k}}\right)^2}+\frac{d+2-\eta_{A,k}}{\left(k^2+\frac{{m}_{A,k}^2}{Z_{A,k}}+2Z_{\phi,k}{e}_k^2{\rho}_0\right)^2}\frac{1}{k^2+\frac{2{\rho}_0{V}_k''({\rho}_0)}{Z_{\phi,k}}}\Bigg]\nonumber\\
&&\hspace{-3.5cm}-2\Omega_d k^{d+2}{e}_k^2\frac{d+2-\eta_k}{d(d+2)}\Bigg[\frac{2N-1}{k^4}+\frac{1}{\left(k^2+\frac{2{\rho}_0{V}_k''({\rho}_0)}{Z_{\phi,k}}\right)^2}\Bigg],
\eea
\end{widetext}
which is evaluated at the minimum point of the potential, ${V}_k'({\rho}_0)=0$. In the symmetric phase, i.e., at ${\rho}_0=0$ (with the assumptions ${m}_{A,k}^2\to 0$, $\eta_k=0$), (\ref{Eq:massflow_final}) becomes 
\bea
k\partial_k {m}_{A,k}^2= -{4N\bar{e}_k^{2}} \varOmega_{d} \frac{d-2}{d(d+2)} k^2, 
\eea
in agreement with Eq. (27) of \cite{Fejos2017}. It is also compatible with (\ref{Eq:mA2}) after taking its $k$-derivative. However, for ${\rho}_0\neq 0$, by comparing (\ref{Eq:massflow_final}) with (\ref{Eq:mA2}) one notes that the mWTI for the gauge mass and the corresponding flow equation do not match. Only when working in the symmetric background with the previous assumptions does one find compatibility between the two approaches.


\bibliographystyle{apsrev}
\bibliography{reference}

\end{document}